\documentstyle[amsfonts,aps,twocolumn]{revtex}

\newcommand{\new}{\newcommand}
\new{\fii}{\varphi}
\new{\Pee}[1]{P[ #1]}
\new{\pee}[1]{\Pee #1}
\new{\vas}{\langle}
\new{\oik}{\rangle} 
\new{\bg}{\begin}
\new{\mfi}{\begin{eqnarray*}}             
\new{\mff}{\end{eqnarray*}}                 
\new{\mfni}{\begin{eqnarray}}             
\new{\mfnf}{\end{eqnarray}}                 
\new{\scal}[2]{\mbox{$\langle{#1},{#2}\rangle$}}
\new{\ip}{\scal}
\new{\no}[1]{\left\|{#1}\right\|}
\new{\ket}[1]{|{#1}\rangle}
\new{\bra}[1]{\langle{#1}|}
\new\tr[1]{{\mathrm{tr}}\bigl[#1\bigr]}       
\new\hi{{\mathcal H}}
\new\room{\ \ \ \ \ \ }
\new{\cuno}{\mathbb C}

\begin{document}


\title{Teleportation of a state in view of the
quantum theory of measurement}
 
\author{P. Busch\thanks{Electronic address: p.busch@maths.hull.ac.uk}}
\address{Department of Mathematics, University of Hull}
\author{G. Cassinelli\thanks{Electronic address: cassinelli@genova.infn.it}}
\address{Dipartimento di Fisica,
Universit\`a di Genova, I.N.F.N., Sezione di Genova, Via Dodecaneso~33,
16146 Genova, Italy}
\author{E. De Vito\thanks{Electronic address: devito@unimo.it}}
\address{Dipartimento di Matematica, Universit\`a di
Modena, via Campi 213/B, 41100 Modena, Italy and I.N.F.N., Sezione di Genova,
Via Dodecaneso~33, 16146 Genova, Italy}
\author{P. Lahti\thanks{Electronic address: pekka.lahti@utu.fi}} 
\address{Department of Physics,
University of Turku, 20014 Turku, Finland}
\author{A. Levrero\thanks{Electronic address: levrero@genova.infn.it}} 
\address{Dipartimento di Fisica, Universit\`a di
Genova, I.N.F.N.,  Sezione di Genova, Via Dodecaneso~33, 16146 Genova,
Italy}

\maketitle
\begin{abstract}
 We give a  description of the teleportation of an
unknown quantum state which takes into account the action of the
measuring device and manifestly avoids any reference to the postulate of
the state vector collapse. 
\end{abstract}


\section{Introduction}\label{intro}

In 1993, Bennett {\em et al.} \cite{Bennett93} introduced the novel idea of teleporting
a state of a quantum system from Alice to Bob using an auxiliary
system shared by Alice and Bob. The first experimental implementations of this idea were
reported by Boschi {\em et al.} \cite{Boschi98}, Bouwmeester {\em et al.}
\cite{Bouw97}, and Furusawa {\em et al.} \cite{Furu98}. In all of these papers
the justification of teleportation is based on three very different
assumptions: the existence of EPR states, the {\em collapse} of the wave
function after the measurement and the exchange of {\em classical
information} from Alice to Bob.

The existence of EPR states follows directly form the linearity of 
quantum mechanics; it was verified in the last two decades by many
experiments, see for example \cite{epr1} and \cite{epr2}. The collapse of the
wave function after a measurement is not a consequence of quantum mechanics,
it is an extra assumption that gives rise to many conceptual problems, for a review see
\cite{qtm2} or the recent paper \cite{Leggett99}. The
exchange of {\em classical} information is a well-known phenomenon in everyday life,
nevertheless, if quantum mechanics is a general description of the world,
as we believe, one has  to be able to explain the teleportation using only a
{\em quantum language}. 

In this Letter we show that teleportation can be achieved by means of two
suitable unitary transformations $U$ and $W$ acting on the states. The
first, $U$, represents a Bell-state premeasurement, carried out
by Alice on her pair of particles 1 and 2, which produces a superposition of
four orthogonal entangled (EPR) states. The second operation, $W$,
represents a measurement that allows Bob to change the state of his particle 3
into an identical copy of the (unknown) input state of particle 1. 
In the experiment of Furusawa {\em et al.}
\cite{Furu98}, the system is a single mode  of an electromagnetic field, $U$ is
realized by a measurement of the 
quadratures performed by Alice and $W$ by a linear displacement of the
quadratures performed by Bob according to the result of the Alice measurement.
Our analysis proves that this experiment constitutes evidence of teleportation
{\em without the need of any extra assumption}. In the other two experiments
cited above the system is the polarization degree of freedom of a light beam,
only the first transformation is performed while the second is omitted. This
leaves the total system in an entangled state, with Bob's photon potentially
being in one of four states with equal probability. Hence one may be
tempted to conclude that "...teleportation is successfully achieved, albeit
only in a quarter of the cases", \cite{Bouw97},\cite{note1}.  However, this
conclusion rests on the problematic projection postulate in that the
mixed state of Bob's photon is interpreted as a classical mixture.
As we will show, there is no evidence of successful teleportation if the
projection postulate is {\em not} assumed.

The use of the collapse postulate in the teleportation scheme was already
criticized  by Motoyashi {\em et al.} \cite{Motoy97}. These authors proposed a
teleportation scheme based on a conservation law -type of an explanation
of the nonlocal EPR correlations.  In view of the difficulties with the so-called
common cause explanations of the EPR problem, see \cite{Bas} for details, we do
not follow here the ideas presented in   \cite{Motoy97}.

To simplify the exposition, we consider two level systems, such as spin $\frac 12$
systems, as
already discussed by Bennett {\em et al.} \cite{Bennett93}. In the next
section we give a description of teleportation using unitary transformations.  

\section{Teleportation}\label{nota}
We consider three spin  $\frac 12$ systems ${\mathcal S}_1$, ${\mathcal S}_2$,
and ${\mathcal S}_3$,
with the associated Hilbert spaces ${\mathcal H}_1={\mathcal H}_2={\mathcal H}_3={\cuno}^2$.
Fix the unit vectors  $\ket{+} = (1,0)$ and $\ket{-}= (0,1)$ in ${\cuno}^2$; 
we write $\ket{\pm}_i\in{\mathcal H}_i$ whenever we wish to emphasize that the
vectors in question are elements of the Hilbert space ${\mathcal H}_i$, $i=1,2,3$.
The Bell basis of ${\cuno}^2\otimes{\cuno}^2$ is defined as
\[
\Psi^{\pm}   = \frac{1}{\sqrt 2}(\ket{+}\ket{-}\pm \ket{-}\ket{+}), \ 
\Phi^{\pm}   =  \frac{1}{\sqrt 2}(\ket{+}\ket{+}\pm \ket{-}\ket{-}),
\]
where the tensor product of vectors is simply written as  $\phi\Psi$.

The system ${\mathcal S}_1$ is initially in a vector state 
$\phi = a\ket{+}_1+b\ket{-}_1$ (with $|a|^2+|b|^2=1$),
which is to be teleported to the system ${\mathcal S}_3$. To achieve this,
${\mathcal S}_2$ and ${\mathcal S}_3$  are prepared initially in the entangled state $\Psi_{23}^- $. 
Then ${\mathcal S}_1+{\mathcal S}_2+{\mathcal S}_3$ is initially in the product state 
$\phi\Psi_{23}^-$.

To teleport the state $\phi$ of ${\mathcal S}_1$ to ${\mathcal S}_3$, the Bennett {\em et
al.} \cite{Bennett93} protocol requires performing a Bell state measurement on
${\mathcal S}_1+{\mathcal S}_2$, that is, a measurement of an observable $A_{12}$
with its simple eigenvectors given by the Bell basis $\Psi^{\pm}_{12}, \Phi^{\pm}_{12}$.

In our analyis of the interaction between the system and the measuring probe in
the context of quantum mechanics we use the notion of {\em premeasurement}
\cite{BCL90,Peres}. 
Let ${\mathcal H}_0$ be the Hilbert space of the measuring probe 
${\mathcal S}_0$. Since the probe must have four orthogonal indicator states corresponding to
the eigenvectors $\Psi^{\pm}_{12}, \Phi^{\pm}_{12}$, one can  assume that
${\mathcal H}_0={\cuno}^4$.
Let $\eta$ be the initial vector state of the probe, and let $\{\eta_1, \eta_2,\eta_3,\eta_4\}$
be a fixed basis  of ${\mathcal H}_0$. One can prove \cite{BCL90} that the
most general measurement interaction between ${\mathcal S}_1+{\mathcal S}_2$ and the probe
is given by a unitary operator 
$U:{\mathcal H}_0\otimes{\mathcal H}_1\otimes{\mathcal H}_2
\to{\mathcal H}_0\otimes{\mathcal H}_1\otimes{\mathcal H}_2$
 of the form
\mfi
U(\eta\Psi^+)  =  \eta_1 \chi_1 & \room & 
U(\eta\Psi^-)  =  \eta_2  \chi_2 \\
U(\eta\Phi^+) = \eta_3  \chi_3  & \room & 
U(\eta\Phi^-) = \eta_4  \chi_4,  
\mff
where $\chi_i$ are four unit vectors in ${\mathcal H}_1\otimes
{\mathcal H}_2$. The choice
$\chi_1=\Psi^+$, $\chi_2=\Psi^-$, $\chi_3=\Phi^+$, $\chi_4=\Phi^-$ corresponds
to a von Neumann-L\"uders measurement.

Adopting this description, the state of the system and the probe after
the measurement is
\begin{eqnarray*}
(U&\otimes& I_3)(\eta\phi\Psi^-_{23})\\ 
&=& \frac 12\eta_1\chi_1\phi_1  + \frac
12\eta_2\chi_2 (\phi_2) + \frac 12\eta_3\chi_3\phi_3+\frac 12\eta_4\chi_4\phi_4,
\end{eqnarray*}
where we have used the notations
\begin{eqnarray*}\label{E:55}
\phi_1 &:=& -a\ket{+}+b\ket{-}, \ 
\phi_2 := -\phi, \\
\phi_3 &:=&  a\ket{-}-b\ket{+}, \ 
\phi_4 :=  a\ket{-}+b\ket{+}.
\end{eqnarray*}
Since the vector state $(U\otimes I_3)(\eta\phi\Psi^-_{23})$ is  entangled, the (reduced)
states of the systems ${\mathcal S}_i$ are not vector states but mixed
states represented by the following density operators:
\mfi
T_0^f &=& \frac 14 P[\eta_1]+\frac 14 P[\eta_2]+\frac 14 P[\eta_3]+\frac 14 P[\eta_4] = \frac 14 I_0,\\
T_3^f &=&  \frac  14P[\phi_1]+\frac  14P[\phi]+  \frac  14P[\phi_3]+
\frac  14P[\phi_4] = \frac 12 I_3,\\
T_{12}^f &=& \frac  14P[\chi_1]+ \frac  14P[\chi_2]+
\frac  14P[\chi_3]+\frac  14P[\chi_4],\\  
T_{012}^f &=& \frac 14\sum_{i,j=1}^4\, \ip{\phi_i}{\phi_j}\, 
\ket{\eta_i\chi_i}\bra{\eta_j\chi_j}\\
T_{123}^f &=&\frac  14P[\chi_1\phi_1] +
\frac  14P[\chi_2\phi] +
\frac  14P[\chi_3\phi_3]
+ \frac  14P[\chi_4\phi_4],
\mff  
where $P[\xi]$ denotes the projection on the one dimensional space generated
by the unit vector $\xi$. We note that the mixed states $T_{0}^f$ and  
$T_{3}^f$ are completely
unpolarized. 

It is important to realize that it is inconsistent to
interpret the state $T_{3}^f$ as a classical mixture
of its component vector states $\phi_1$, $\phi_2=-\phi$, $\phi_3$, $\phi_4$.
Such an {\em
ignorance interpretation} is incompatible with the pure nature of the
entangled total state $(U\otimes I_3)(\eta\phi\Psi^-_{23})$ of system
plus probe. Although this crucial point has been known for many years,
see, e.g., the classic analysis of d'Espagnat \cite{Esp71}, it is frequently
ignored in the discussion of experiments involving entanglement (for a
notable recent exception in the context of teleportation, 
see \cite{KB00}). In fact
the EPR experiment itself can be taken as a demonstration of the
inadmissability of the ignorance interpretation for the reduced mixed
states, cf., e.g., \cite{qtm2}. Formally, this phenomenon is reflected
in the fact that the decomposition of a mixed state into
vector (or pure) states is never unique (for a full analysis, see
e.g. \cite{GEA97}). 

The same remarks apply to the reduced state $T_0^f$ of the probe,
leading to the objectification problem, that is, the impossibility of
explaining {\em within quantum mechanics} the occurrence of definite
outcomes as a result of the measurement, represented by the appearance 
of one of the possible indicator states $\eta_i$. As is generally
acknowledged, this problem constitutes a serious challenge to the view
of quantum mechanics as a consistent and universal theory. The commonly
accepted, pragmatic way of dealing with it is the application of the
collapse postulate. While this seems good for (almost) all practical
purposes in the discussion of
most experiments, we hope the above remarks serve as a reminder of the
fundamental inconsistency of such an approach.

It follows that after the premeasurement $U$ the state of ${\mathcal S}_3$
is $T_3^f$, and this state yields no evidence of quantum
teleportation as can be seen by computing the fidelity, 
\[
{\rm tr}\left(P[\phi]T_3^f\right)=\frac 12.
\]
It is only when the collapse is postulated that one can conclude that 
``teleportation is successfully achieved ... in a quarter of the cases."

However, if the Bell state premeasurement $U$ is followed by another unitary
operation $W$, which models the classical communication and ensuing
manipulation of ${\mathcal S}_3$, then the collapse postulate can be
avoided, thus bypassing the objectification problem.

This final step consists of a rotation of the state of system 
${\mathcal S}_3$ depending on the result of the previous measurement. 
Since the probe is itself a quantum system, we can model this conditional
manipulation based on a classical information transfer as another 
premeasurement performed on the system
${\mathcal S}_0+{\mathcal S}_3$ by means of a
unitary operator 
$W:{\mathcal H}_0\otimes{\mathcal H}_3\to
{\mathcal H}_0\otimes{\mathcal H}_3$, 
\mfi
W(\eta_1\ket{+}) = -\eta_1\ket{+}& \room & 
W(\eta_1\ket{-}) =  \eta_1\ket{-} \\
W(\eta_2\ket{+}) = -\eta_2\ket{+}&  \room & 
W(\eta_2\ket{-}) = -\eta_2\ket{-}\\ 
W(\eta_3\ket{+}) = -\eta_3\ket{-}& \room & 
W(\eta_3\ket{-}) = \eta_3\ket{+}\\ 
W(\eta_4\ket{+}) =\ \,\, \eta_4\ket{-}& \room & 
W(\eta_4\ket{-}) = \eta_4\ket{+}
\mff
In doing so, there is no need to account for the actual occurrence of a
result, and the objectification problem is circumvented.
Regarding $W$ as a linear transformation  on 
${\mathcal H}_0\otimes{\mathcal H}_1\otimes{\mathcal H}_2\otimes{\mathcal H}_3 $
in a natural way, one obtains the final state
$$W\left( U(\eta\phi\Psi^-) \right) =
\frac 12(\eta_1\chi_1+\eta_2\chi_2+\eta_3\chi_3+\eta_4\chi_4)\phi.$$
The (reduced) state of the system ${\mathcal S}_3$ is now the vector state
$\phi$, which has been teleported with 100$\%$ efficiency 
from the system ${\mathcal S}_1$ to ${\mathcal S}_3$.

We note that since $U$ and $W$
are unitary operators, they can be realized (at least theoretically) by suitable quantum
Hamiltonians.

In order to highlight the difference between our approach and the ones based on the
projection postulate, we modify slightly the form of $W$. We define 
a unitary transformation $W_\theta$ (with $\theta\in [0,2\pi)$) in such a way that
it is equal to $W$ save that
\[
W_\theta(\eta_2\ket{+}) =- e^{i\theta}\eta_2\ket{+}.
\]
A simple calculation shows that 
\begin{eqnarray*}
W_\theta\left( U(\eta\phi\Psi^-) \right) &=&
\frac 12(\eta_1\chi_1+\eta_3\chi_3+\eta_4\chi_4)\phi
+ \frac 12 \eta_2\chi_2\phi_{\theta},\\
\phi_\theta&=&(e^{i\theta} a \ket{+} + b \ket{-}),
\end{eqnarray*}
and the final reduced state of ${\mathcal S}_3$ is now
\[
T^f_3(\theta)= \frac 34 P[\phi] + \frac 14 P[\phi_{\theta}],
\]
If $\theta\neq 0$, $a\ne 0$,  and $b\neq 0$, 
then $T^f_3(\theta)$ is a mixed state different from the pure state $P[\phi]$, 
so that,  without the collapse assumption or ignorance interpretation, 
there is no teleportation. 
Nevertheless, in the limit $\theta\to 0$, we have $\phi_\theta\to\phi$,
and so $T^f_3(\theta)$ 
continuously approaches $P[\phi]$, thus leading to perfect teleportation
upon reaching $\theta=0$.
By contrast, if one
accepts  the projection postulate, one could say that the probe collapses into
one of the four states $\eta_1,\ldots,\eta_4$, each of them corresponding to a definite
value of the measurement, and, performing a coincidence detection, one could say that
${\mathcal S}_3$ is found in the exact target state
$\phi$ in $75\%$ of the cases. In doing so one has to face the fact that, in the limit
$\theta\mapsto 0$, the rate of {\em correct} target states [successful perfect teleportations]
jumps from $75\%$ to $100\%$; this fact contrasts with the common understanding that 
Nature is {\em continuous} with respect the variation of {\em parameters} such as
$\theta$. It is interesting to observe that the teleportation fidelity,
being a statistical measure, cannot distinguish between the two
interpretations of $T_3^f(\theta)$ (a) as a quantum mixed state -- in
which case one can speak of {\em approximate} teleportation as
$\theta\to 0$, and (b) as a classical mixture -- in which case there is
perfect teleportation in 75$\%$ of the cases. In either case, the
fidelity is ${\rm tr}\left(P[\phi]T_3^f(\theta)\right)$. Nevertheless
this does not justify the conclusion that the two approaches are
physically equivalent.

Finally we note that, in 
our approach, the coincidence detection can be simply interpreted as  a measurement
of the observable (with possible outcomes $0$ and $1$) on ${\mathcal H}_0\otimes {\mathcal H}_3$
\[
A=P[\eta_1\phi] +P[\eta_3\phi]+ P[\eta_4\phi].
\]
Indeed, if we regard $A$ in a natural way as operator on 
${\mathcal H}_0\otimes{\mathcal H}_1\otimes{\mathcal H}_2\otimes{\mathcal H}_3 $,
we have that 
\[\vas W_\theta\left( U(\eta\phi\Psi^-) \right) |A|W_\theta\left(
  U(\eta\phi\Psi^-) \right)\oik = \frac 34.
\]

\section{Conclusion}\label{con}
 We have given a measurement theoretical description of the teleportation of an
unknown quantum state which takes into account the action of the
measuring device and manifestly avoids any reference to the state vector collapse
postulate.  
The teleportation  process can be
described as a measurement whose outcome is certain; equivalently the process is
realized by the action of a unitary map. In other words, teleportation can be
carried out as a deterministic process and hence does not require stochastic
(or quantum) jumps in the sense expressed by the collapse postulate.

\end{document}